\documentclass[%
 aip,
 amsmath,amssymb,
 reprint,%
]{revtex4-1}

\usepackage{graphicx}
\usepackage{dcolumn}
\usepackage{bm}

\usepackage[utf8]{inputenc}
\usepackage[T1]{fontenc}
\usepackage{mathptmx}
\usepackage{etoolbox}
\usepackage{xcolor}

\def\bv{{\mathbf{v}}}
\def\bx{{\mathbf{x}}}
\def\dx{d\bx}
\def\br{{\mathbf{r}}}
\def\bj{{\mathbf{j}}}
\def\dr{d\br}

\makeatletter
\def\@email#1#2{%
 \endgroup
 \patchcmd{\titleblock@produce}
  {\frontmatter@RRAPformat}
  {\frontmatter@RRAPformat{\produce@RRAP{*#1\href{mailto:#2}{#2}}}\frontmatter@RRAPformat}
  {}{}
}%
\makeatother
\begin{document}

\preprint{AIP/123-QED}

\title{Variational formulation for the dynamics of soft matter including inertia}

\author{Andrew J.~Archer}
\email{a.j.archer@lboro.ac.uk}
\affiliation{Department of Mathematical Sciences and Interdisciplinary Centre for Mathematical Modelling, Loughborough University, Loughborough, Leicestershire, LE11 3TU, United Kingdom}

\date{\today}

\begin{abstract}
The motion of liquids and soft matter is over-damped and ‘slow’ when viscosity dominates. In this (low Reynolds-number) limit and when the system is isothermal, the equations of motion may be generated via Onsager’s variational principle, which neglects inertia. This variational approach is immensely powerful, being used to obtain equations of motion for colloidal fluids, droplets on surfaces and much more. However, inertia can play a role, manifesting as vibrations and under-damped motion. Here we show how to extend this variational framework so that it remains valid for when damping/dissipation and inertia are both equally important.
\end{abstract}

\maketitle


{\em Introduction:}
In physics there is a long history of formulating dynamical equations as functional minimisation problems.
Variational principles are elegant and powerful formulations of the governing physics.
For over-damped dissipative systems, where frictional forces dominate, Onsager's variational principle (OVP) is a minimum free energy dissipation principle that allows us to express the governing equations in a variational manner.\cite{onsager1931reciprocal_1, groot1964non}
This approach directly connects to the extensive knowledge we have from thermodynamics, statistical mechanics, Landau theory and its symmetry based arguments, which all lead to good approximations for the free energy, $F$.\cite{chaikin2000principles}
OVP takes us directly from the free energy to the time evolution equations.
The OVP approach is thus extremely powerful, but it is only applicable in the over-damped (slow dynamics) limit.
Here, we extend the OVP approach to include under-damped dynamics and other inertial effects, yielding a theoretical framework for generating the equations of motion
of liquids and other soft matter systems whenever damping/dissipation and inertia are equally relevant.

The OVP approach has been successfully used in recent years to derive equations of motion for fluids and other soft condensed matter systems in the viscosity dominated (low Reynolds-number) regime.\cite{doi2011onsager, doi2013soft, doi2019application, wang2021onsager}
For colloidal fluids, the particle equations of motion can often be approximated as a Brownian motion, i.e.\ via over-damped stochastic equations of motion.
For such systems, Brader and Schmidt \cite{schmidt2013power, schmidt2022power} developed a formally exact theory which they named power functional theory (PFT). Their theory has exactly the same form as OVP and generates the dynamical equations for the time evolution of the ensemble (noise) averaged local density $\rho(\br,t)$ at position $\br$ and time $t$.
PFT shows that the OVP formulation of the dynamics is exact, albeit not all the involved quantities are known exactly -- we say more about this below.

The starting point for the OVP approach is an expression for the free energy $F$ that incorporates the relevant physics, written as a function/functional of the slow variables/fields characterising the state of the system.
For example, for colloidal suspensions, the relevant field is the local density $\rho(\br,t)$ at position $\br$ and time $t$.
Or, for liquid drops on surfaces, the relevant quantity is the thickness of the liquid $h(\bx,t)$ over position $\bx$ on the surface.\cite{oron1997long, degennes2003capillarity, craster2009dynamics}
For each relevant variable/field there is an associated velocity, $\bv(\br,t)$.
Next, one constructs the Rayleighan
\begin{equation}
\label{eq:1}
R = \dot{F}+\Phi,
\end{equation}
where $\dot{F}$ denotes the time derivative of $F$, and $\Phi$ is the dissipation function/functional.
The exact form of $\Phi$ is generally unknown, as evident in the PFT formulation for colloidal particles.\cite{schmidt2013power, schmidt2022power, lutsko2021reconsidering}
However, the leading order contributions must be quadratic in the velocity $\bv$, and in many systems using just this is a good approximation.\cite{doi2011onsager, doi2013soft, doi2019application}
{This is because the dissipation $\Phi\geq0$ must be positive definite, and (as we illustrate below) if there are any terms in $R$ that are linear in $\bv$, this would correspond to a work power done by an {\em active} force, i.e.\ to the particles being {\em active} particles.\cite{wang2021onsager} We assume henceforth that this is not the case.}
The dynamical equations are then given by minimising the Rayleighan $R$ with respect to the relevant velocities/currents, giving the stationary condition(s)
\begin{equation}
\label{eq:2}
\frac{\delta R}{\delta \bv}=0,
\end{equation}
i.e.\ the functional derivative of $R$ with respect to the velocities $\bv$ must equal zero.
The structure of Eq.~\eqref{eq:2} already indicates it applies only in the over-damped limit.
Since it involves just the velocity $\bv$ and not the acceleration $\dot{\bv}$, it therefore neglects inertia, preventing accurate description of under-dampened dynamics.
Note that if $R$ is approximated by a function (not a functional) of $\bv$, then \eqref{eq:2} should be interpreted as just a partial derivative.\cite{doi2011onsager, doi2013soft, doi2019application}

From \eqref{eq:2} we obtain an expression for the velocity $\bv$ in terms of (functional) derivatives of $F$.
For colloidal fluids (see below), this expression can be used with the continuity equation,
\begin{equation}
    \label{eq:continuity}
    \frac{\partial \rho}{\partial t}=-\nabla\cdot(\rho \bv),
\end{equation}
to obtain the dynamics.
If the relevant slow field is not conserved or there is a non-conserved aspect to the dynamics, then instead we have $\frac{\partial h}{\partial t}=-\nabla\cdot(h \bv)+\sigma$.
This includes the non-conserved (source or sink) term $\sigma$. Such a term arises e.g.\ in the case of liquids drops on surfaces when there is evaporation from the drop or condensation of vapour in the air onto the drop.\cite{oron1997long, MoHo1980jcis, Ajae2005pre, thiele2009modelling, thiele2010thin,toth2026variational}
More broadly, the source term $\sigma$ may describe growth processes in living matter,\cite{trinschek2017continuous} chemical reactions in mixtures\cite{voss2025chemomechanical, niehoff2026gradient} and all sorts of other physical mechanisms leading to creation or destruction of the substance.
We hope that it is now becoming apparent that this variational approach is extremely powerful.

This brings us to our central hypothesis: for systems where inertial effects are as equally relevant as viscous/dissipative effects, we argue that Eq.~\eqref{eq:2} should be replaced with
\begin{equation}
\label{eq:4}
\frac{\delta R}{\delta \bv}=-\frac{\partial}{\partial t}\left(\frac{\delta K}{\delta \bv}\right),
\end{equation}
where $K$ is the kinetic energy of the system, which is a function(al) of $\bv$.


{\em Dynamics of a single particle:}
Rather than starting with a derivation of Eq.~\eqref{eq:4} (which we postpone to later), for now we assume Eq.~\eqref{eq:4} and explore where that assumption leads to.
As alluded to above, we are building towards showing that Eq.~\eqref{eq:4} can be used to formulate the dynamical equations for the (number) density distribution $\rho(\br,t)$ of $N$ interacting colloidal particles suspended in a fluid.
However, as an instructive `warm-up problem' it helps if we first consider the much simpler case of a single colloid that is suspended in a background fluid and under the influence of an external potential $U(\br)$.
In this case, the free energy is simply the potential energy, i.e.\ $F=U(\br)$.
Differentiating with respect to time, we obtain $\dot{F}=\nabla U
\cdot\bv$ (chain rule), where $\bv = \dot{\br}$ is the velocity and $\nabla=\partial/\partial \br$.

As mentioned, the leading order contribution to $\Phi$ must be quadratic in $\bv$, so we write $\Phi=\frac12\gamma\bv^2+\Phi_{ex}$, where $\gamma$ is a friction constant, the value of which will become clear shortly, and $\Phi_{ex}$ is the remainder, incorporating everything neglected in the leading order quadratic contribution.
Assuming $\Phi_{ex}\approx0$ and then plugging these into Eq.~\eqref{eq:1} gives, $R=\nabla U \cdot\bv+\frac12\gamma\bv^2$.
As a first check, we substitute this into Eq.~\eqref{eq:2}, i.e.\ we differentiate the Rayleighan $R$ with respect to $\bv$ (keeping $\br$ constant), to obtain
\begin{equation}\label{eq:5}
    \gamma \bv +\nabla U=0.
\end{equation}
This is simply the over-damped dynamical equation for a particle in a fluid, given as the balance of the Stokes drag force and the force due to the potential $U$, as long as we identify $\gamma=6\pi\eta {\cal R}$ (the Stokes law), where $\eta$ is the fluid viscosity and ${\cal R}$ is the radius of the particle.
We can also now see that if we had included a linear term in the dissipation, $\Phi=\mathbf{a}\cdot\mathbf{v}+\frac12\gamma\bv^2+\Phi_{ex}$, then instead of Eq.~\eqref{eq:5} we would have ended up with the equation of motion $\gamma \bv =\mathbf{a}-\nabla U$, which is a simple model for active matter, where $\mathbf{a}/\gamma$ is the self-propulsion speed.

In the under-damped case, we must consider the kinetic energy $K=\frac12m\bv^2$, where $m$ is the mass of the particle. Plugging this into Eq.~\eqref{eq:4}, together with the result in Eq.~\eqref{eq:5} for the left hand side, gives
\begin{equation}\label{eq:6}
    -\gamma \bv -\nabla U=m\dot{\bv},
\end{equation}
which of course is just Newton's equations for a particle under the influence of the Stokes drag force and an external force $-\nabla U$.
We are now ready to tackle the real problem of interest.

{\em Under-damped interacting colloids:}
We now consider systems of $N$ interacting Brownian particles with under-damped (Langevin) equations of motion.
In the over-damped limit, Eq.~\eqref{eq:2} is exact and is the central result of PFT, which also gives precise statistical mechanical definitions of the quantities in $R$.\cite{schmidt2013power, schmidt2022power}
In the under-damped case, formally exact isothermal equations of motion for the coupled fields $\rho$ and $\bv$ that were obtained in Ref.~\onlinecite{archer2009dynamical}.
We now show here that these equations can instead be obtained from Eq.~\eqref{eq:4}.
In other words, Eq.~\eqref{eq:4} is the generalised variational principle (or PFT) for generating the dynamical equations.
These are also often referred to as `dynamical density functional theory' (DDFT),\cite{marconi1999dynamic, archer2004dynamical, archer2009dynamical, hansen2013theory, te2020classical} building on the extremely powerful equilibrium classical density functional theory (DFT).\cite{hansen2013theory, evans1979nature}
DFT is a theory for the free energy functional $F[\rho]$.
The advantage of DDFT/PFT is that it builds on this extensive body of knowledge.

The first term in the Rayleighan \eqref{eq:1} can be evaluated as follows
\begin{align}
    \dot{F}=&\int\frac{\delta F}{\delta \rho}\frac{\partial \rho}{\partial t} \dr\nonumber\\
    =&-\int\frac{\delta F}{\delta \rho}\nabla\cdot(\rho\bv) \dr\nonumber\\
    =&\int\rho\bv\cdot\nabla\frac{\delta F}{\delta \rho} \dr,
    \label{eq:7}
\end{align}
where the first line is just the chain rule for functionals, the second line comes from using Eq.~\eqref{eq:continuity} and the last from an integration by parts and also assuming that the boundary terms are zero, i.e.\ that either the flux $\bj=\rho\bv$ is zero on the boundaries or that there are periodic boundary conditions.
The dissipation functional can be written as \cite{schmidt2022power}
\begin{equation}
    \Phi=\int \frac{\gamma}{2} m\rho \bv^2 \dr+\Phi_{ex},
\end{equation}
where $\gamma$ is the same friction coefficient introduced above Eq.~\eqref{eq:5}.
Notice the leading order term is quadratic in $\bv$, while all higher order contributions are incorporated in $\Phi_{ex}$.
Taking the functional derivative of the Rayleighan $R$ with respect to $\bv$ (keeping $\rho$ fixed), we obtain
\begin{equation}\label{eq:9}
    \frac{\delta R}{\delta \bv}=\rho\nabla\frac{\delta F}{\delta \rho}+\gamma m \rho \bv+\frac{\delta \Phi_{ex}}{\delta \bv}.
\end{equation}
In the over-damped case \eqref{eq:2}, where the expression above equals zero, together with Eq.~\eqref{eq:continuity}, this is just the PFT of Schmidt and Brader.\cite{schmidt2013power, schmidt2022power}
Making the further approximation that $\Phi_{ex}=0$, then in the over-damped limit we obtain that the flux $\rho\bv=-\Gamma\rho\nabla\delta F/\delta\rho$, where $\Gamma=1/(m\gamma)$, which can be inserted into Eq.~\eqref{eq:continuity} to obtain the usual adiabatic DDFT equation\cite{marconi1999dynamic, archer2004dynamical, hansen2013theory, te2020classical}
\begin{equation}
{\frac{\partial \rho}{\partial t}=\nabla\cdot\left[\Gamma\rho\nabla\frac{\delta F}{\delta \rho}\right].}
\end{equation}

Returning to the under-damped case, we must consider the kinetic energy
\begin{equation}\label{eq:10}
    K=\int\frac12m\rho\bv^2\dr.
\end{equation}
From this we obtain
\begin{equation}\label{eq:11}
    \frac{\partial}{\partial t}\left(\frac{\delta K}{\delta \bv}\right)=\frac{\partial}{\partial t}\left(m\rho\bv\right)=m\frac{\partial\bj}{\partial t}.
\end{equation}
On plugging the results in Eqs.~\eqref{eq:9} and \eqref{eq:11} into Eq.~\eqref{eq:4} and dividing through by $m$, we obtain
\begin{equation}\label{eq:12}
    \frac{\partial\bj}{\partial t}+\gamma\bj+\frac{1}{m}\frac{\delta \Phi_{ex}}{\delta \bv}+\frac1m\rho\nabla\frac{\delta F}{\delta \rho}=0.
\end{equation}
This is identical to Eq.~(20) in Ref.~\onlinecite{archer2009dynamical}, although in the notation of \onlinecite{archer2009dynamical} we would instead write the term $\frac{1}{m}\frac{\delta \Phi_{ex}}{\delta \bv}=A(\br,t)$.
The starting point of the derivation of this equation in Ref.~\onlinecite{archer2009dynamical} is the $N$-particle Kramers (Fokker-Plank) equation.
The approach of Ref.~\onlinecite{archer2009dynamical} is to integrate Kramers' equation with respect to $(N-1)$ degrees of freedom, in order to obtain the time evolution equation for the one-body density $\rho(\br,t)$.
It should also be emphasised that the free energy $F$ in Eq.~\eqref{eq:12} is a quantity given in Ref.~\onlinecite{archer2009dynamical} in terms of non-equilibrium distribution functions, which we approximate by corresponding equilibrium distributions.
Another way to say this is that $F$ in Eq.~\eqref{eq:12} is a
non-equilibrium free energy, which we then approximate by the corresponding equilibrium free energy.
Underpinning this argument is the fact that the mapping from the density to a time-dependent external potential is unique.
This and the required explicit conditions (such as no-flux at the boundaries) are derived in Ref.~\onlinecite{klatt2026foundation}.

The main conclusion we draw from the derivation above is that the formally exact dynamical equation \eqref{eq:12} is {\em generated} by the expression in Eq.~\eqref{eq:4}.
Thus, we may view \eqref{eq:4} as the PFT or generating principle for the dynamics in the under-damped isothermal case.
We must remember of course that for most systems in practice neither $F[\rho]$ nor $\Phi_{ex}[\rho,\bv]$ are known exactly, as discussed in Ref.~\onlinecite{schmidt2022power} in the context of the over-damped case.

In the over-damped case, there has been good progress to develop suitable approximations for $\Phi_{ex}$, that is summarised well in Ref.~\onlinecite{schmidt2022power}, where $\Phi_{ex}$ is referred to as the excess power functional $P^{exc}$.
The derivative $-\delta P^{exc}/\delta \bv$ gives the `super-adiabatic forces' that have been studied extensively by Schmidt and coworkers -- see\cite{schmidt2013power, schmidt2022power, stuhlmuller2018structural, treffenstadt2020memory, deLasHeras2020flow, geigenfeind2020superadiabatic, treffenstadt2022dynamic, deLasHeras2023perspective}, for further details.
For example, the following has been proposed\cite{stuhlmuller2018structural, schmidt2022power}
\begin{equation}\label{eq:P_ex_visc}
{\Phi_{ex}\approx\int\left[\eta(\nabla\times\bv)^2+\zeta(\nabla\cdot\bv)^2\right]\dx,}
\end{equation}
with shear and bulk viscosities $\eta=\eta(\rho)$ and $\zeta=\zeta(\rho)$, and also expressions for $\Phi_{ex}$ involving time integrals and memory effects.
We see no reason why all the approximations developed for the over-damped case should not also be applicable in the under-damped case.
In Ref.~\onlinecite{archer2009dynamical} a simple local-equilibrium approximation leads to $A(\br,t)\approx A_{le}(\br,t)=\nabla\cdot(\rho \bv \otimes \bv)$, where $\otimes$ denotes a dyadic product, which then results in Eq.~\eqref{eq:12} having the form of a generalised Stokes equation.
Clearly, further work applying and perhaps developing other approximations for $\Phi_{ex}$ is required, building on the work in\cite{schmidt2013power, schmidt2022power, stuhlmuller2018structural, treffenstadt2020memory, deLasHeras2020flow, geigenfeind2020superadiabatic, treffenstadt2022dynamic, deLasHeras2023perspective}.
We hope the ideas presented here provide impetus for such activity.

Having demonstrated that Eq.~\eqref{eq:4} generates the correct equations of motion for both a single particle in a fluid and also the equations of motion for the density distribution of interacting Brownian colloidal particles, we feel we have sufficient evidence to postulate that Eq.~\eqref{eq:4} is in fact much more general.
In other words, whenever inertial effects are expected to be just as relevant as viscous dissipation, we propose that Eq.~\eqref{eq:4} may be applied entirely in the same spirit as the OVP approach based on Eq.~\eqref{eq:2}.
As discussed in the introduction, the essence of the OVP approach is to identify the relevant slow variables/fields and then express the free energy and Rayleighan in terms of these. 
This view is supported by the arguments of Ref.~\onlinecite{laurila2006interface}, which shows that when a high-dimensional system, with dynamics that can be written as the variation of a Rayleighan $R$, is projected down onto a lower dimensional phase space, then the variational structure is preserved. In other words, the projected dynamics can still be obtained from a stationary condition on a more course-grained Rayleighan $R$.
Such arguments imply that such a coarse graining from microscopic to more mesoscopic degrees of freedom should retain the variational structure implicit in Eq.~\eqref{eq:4}.
Taking this view, we now apply our approach in order to derive equations of motion for liquid films and droplets on surfaces.


{\em {liquid films and drops on solid surfaces:}}
The time evolution equation for the height $h(\bx,t)$ of liquid films/droplets on surfaces (generally referred to as the `thin-film equation') is normally derived via a long-wave analysis of the Navier-Stokes equations,\cite{oron1997long, degennes2003capillarity, craster2009dynamics} although it can instead be derived via the OVP in Eq.~\eqref{eq:2}, as discussed in\cite{qian2006variational, xu2015variational, xu2016variational, Thie2018csa, lopes2018multiple, peschka2018variational, toth2026variational}.
The gradient dynamics formulation of thin-film hydrodynamics \cite{mitlin1993dewetting, thiele2010thin, thiele2011note} helps to see the connections between these two different approaches.
However, there are numerous situations where liquids on surfaces exhibit under-damped vibrational (oscillatory) behaviour, that are not described by the (over-damped) thin-film equation.
By starting from Eq.~\eqref{eq:4}, inertial effects are incorporated, giving a theory that should be capable of describing these cases.

The derivation proceeds in a manner somewhat analogous to the colloidal case in Eqs.~\eqref{eq:7}--\eqref{eq:12}.
Exactly equivalent to Eq.~\eqref{eq:7}, we obtain
\begin{equation}\label{eq:13}
    \dot{F}=\int h\bv\cdot\nabla\frac{\delta F}{\delta h} \dx.
\end{equation}
Note that in the above $h\bv=\bj$, the two dimensional (2D) height-averaged flux over the surface and now $\nabla=\partial/\partial\bx$, where $\bx$ is the 2D position coordinate on the surface.
The dissipation functional is \cite{toth2026variational}
\begin{align}\label{eq:14}
    \Phi=&\int \frac{\bj^2}{2M}\dx+\Phi_{ex}\nonumber \\
    =&\int \frac{3\eta\bv^2}{2h}\dx+\Phi_{ex},
\end{align}
where we have used the mobility $M=h^3/(3\eta)$, where $\eta$ is the fluid viscosity, which is derived by assuming no-slip plane Poiseuille film flow over the surface.\cite{oron1997long, degennes2003capillarity, craster2009dynamics, toth2026variational}
From Eqs.~\eqref{eq:13} and \eqref{eq:14} we obtain
\begin{equation}\label{eq:15}
    \frac{\delta R}{\delta \bv}=h\nabla\frac{\delta F}{\delta h}+\frac{3\eta\bv}{h}+\frac{\delta \Phi_{ex}}{\delta \bv}.
\end{equation}
In the case where the dynamics is over-damped (the usual thin-film long-wave limit) then the right hand side of \eqref{eq:15} equals zero.
If we also assume that $\Phi_{ex}=0$ and that evaporation $\sigma$ can be neglected, then Eq.~\eqref{eq:15} gives the over-damped result
\begin{equation}\label{eq:16}
    \bj=h\bv=-\frac{h^3}{3\eta}\nabla\frac{\delta F}{\delta h}.
\end{equation}
The usual thin-film approximation for the free energy is\cite{oron1997long, degennes2003capillarity, craster2009dynamics}
\begin{equation}\label{eq:free_thin}
    F[h]=\int\left(f(h)+\frac{\gamma}{2}(\nabla h)^2\right)\dx,
\end{equation}
where $f(h)$ is the binding potential and $\gamma$ is the surface tension.
Taking this free energy, together with the expression for the flux in Eq.~\eqref{eq:16} and the continuity equation gives the usual thin-film equation\cite{oron1997long, degennes2003capillarity, craster2009dynamics,qian2006variational, xu2015variational, xu2016variational, Thie2018csa, lopes2018multiple, peschka2018variational, toth2026variational}
\begin{equation}
\frac{\partial h}{\partial t} =\nabla\cdot\left[\frac{h^3}{3\eta}\nabla\left(\Pi(h)+\gamma\nabla^2h\right)
\right],
\end{equation}
where the disjoining pressure $\Pi(h)=-\partial f(h)/\partial h$.

In the under-damped case, we must consider the kinetic energy
\begin{equation}\label{eq:film_KE}
    K=\int\frac{3\varrho_m\,\bj^2}{5h}\dx=\int\frac35h\varrho_m\,\bv^2\dx,
\end{equation}
where $\varrho_m$ is the mass density of the liquid (assumed constant within the liquid film).
The above result comes from assuming that the velocity profile $u(\br,t)$ within the liquid film is a plane Poiseuille film flow with no slip at the solid surface, i.e.\ in one dimension having the parabolic form $u(z)=u_0(z^2/2-hz)$ for $0<z<h$, where $z$ is the distance perpendicular to the surface.
The corresponding flux is $j=-\int_0^hu(z)dz$ (see e.g.\cite{oron1997long,toth2026variational}) and the kinetic energy density is therefore $\int_0^h\frac12\varrho_m u(z)^2dz=\frac{3\varrho_m\,j^2}{5h}$.

Plugging Eq.~\eqref{eq:film_KE} into the right hand side of Eq.~\eqref{eq:4} and using the result in \eqref{eq:15} for the left hand side, we obtain
\begin{equation}\label{eq:18}
    h\nabla\frac{\delta F}{\delta h}+\frac{3\eta\bj}{h^2}+\frac{\delta \Phi_{ex}}{\delta \bv}=-\frac65\varrho_m\frac{d\bj}{dt}.
\end{equation}
Taking this result together with the approximation for the free energy in Eq.~\eqref{eq:free_thin} and also assuming that the binding potential $f(h)\approx0$ and so may be neglected, we obtain
\begin{equation}\label{eq:20}
    \frac{d\bj}{dt}=\frac{5h}{6\varrho_m}\nabla (\gamma\nabla^2h)-\frac{5\eta}{2h^2\varrho_m}\bj - \frac{5}{6\varrho_m}\frac{\delta \Phi_{ex}}{\delta {\bv}} .
\end{equation}
Comparing the above result with Eq.~(8) in Ref.~\onlinecite{ruyer2002further} allows to see what would be neglected if one were to assume $\Phi_{ex}=0$, and already gives hints for how to choose $\Phi_{ex}$, as do PFT results, such as Eq.~\eqref{eq:P_ex_visc}.
In other words, Eq.~\eqref{eq:20} together with a suitable approximation for $\Phi_{ex}$ is a good place to start when deriving thin-film flow type equations incorporating inertia.

A fruitful future avenue for research that we do not pursue here is to explore what choice of approximation for $\Phi_{ex}$ leads to either (i) the Benney equation,\cite{benney1966long} (ii) the thin-film theory of Ref.~\onlinecite{zitz2019lattice}, which is solved using lattice-Boltzmann methods, (iii) the weighted residuals theory, \cite{ruyer2000improved, ruyer2002further, holroyd2024linear} or possibly an improvement on these.
We believe such an exploration will be extremely fruitful.
We expect that generating evolution equations for $h(\br,t)$ in this way will lead to identifying new physics and specifically to good approximations for $\Phi_{ex}$.

Before we move on, we note that setting $\Phi_{ex}=0$ and introducing the integrating factor $\phi=\exp(\frac{5\eta}{2\varrho_m}\int\frac{1}{h^2}dt)$ allows us to rewrite Eq.~\eqref{eq:18} as
\begin{equation}
    \frac{d}{dt}(\phi\bj)=
    -\frac{5\phi h}{6\varrho_m}\nabla\frac{\delta F}{\delta h},
\end{equation}
which can be integrated to give
\begin{equation}
    \phi(h,t)\bj=
    -\int_0^t\frac{5\phi(h,t') h}{6\varrho_m}\nabla\frac{\delta F}{\delta h}dt',
\end{equation}
or equivalently
\begin{equation}\label{eq:23}
    \bj=
    -\int_0^t{\cal M}\nabla\frac{\delta F}{\delta h}dt',
\end{equation}
with the following explicit result for the memory function: ${\cal M}=5h'\phi'/(6\varrho_m\phi)$, where quantities with a prime are evaluated at time $t'$, while the $\phi$ without a prime is evaluated at the later time $t$.s
The result in Eq.~\eqref{eq:23} is precisely what one would expect if one were to derive the thin-film equation via the Mori-Zwanzig formalism.\cite{te2024microscopic}
The integrating factor approach used here can also be applied to Eq.~\eqref{eq:12} in the colloidal case when $\Phi_{ex}=0$.
The resulting memory function in that case is a simple exponential. See also Refs.~\onlinecite{anero2013functional, wittkowski2013microscopic, thewes2024mobility, santra2026exact} for further discussion of such DDFT-type theories that include memory.


{\em A derivation of equation \eqref{eq:4}:}
Having taken a `try it and see' approach to presenting our arguments for why Eq.~\eqref{eq:4} should be taken as the generalisation of OVP to situations where inertial effect must also be included, we now present a derivation of Eq.~\eqref{eq:4} that starts from ideas in particle mechanics.
We commence this by recapitulating some key ideas given in Refs.~\onlinecite{edwards1974theory, whittacker1937}; see also Ref.~\onlinecite{giga2018variational}.
In conservative dynamical systems, the dynamics can be obtained from minimising the time integral of the Lagrangian ${\cal L}$,
\begin{equation}
    \delta \int {\cal L}[\br(t),\dot{\br}(t),\cdots] dt =0,
\end{equation}
which then yields the familiar Euler-Lagrange equations
\begin{equation}\label{eq:EL_eq_1}
    \frac{\delta{\cal L}}{\delta \br} - \frac{d}{dt}\left(\frac{\delta{\cal L}}{\delta \dot{\br}}\right) + \frac{d^2}{dt^2}\left(\frac{\delta{\cal L}}{\delta \ddot{\br}}\right) +\cdots =0.
\end{equation}
To extend to the case where friction is present, we must introduce the dissipation by including another term that depends on the velocities, which to be consistent with previous discussion we denote as $-\Phi(\bv)$, so that the variational principle becomes
\begin{equation}
    \delta \int {\cal L}[\br(t),\dot{\br}(t),\cdots] dt -  \delta \int \Phi[\bv(t),\dot{\bv}(t),\cdots] dt= 0,
\end{equation}
and so Eq.~\eqref{eq:EL_eq_1} becomes
\begin{align}
    \frac{\delta{\cal L}}{\delta \br} - &\frac{d}{dt}\left(\frac{\delta{\cal L}}{\delta \dot{\br}}\right) +\dots
    \nonumber \\
    &-\left[\frac{\delta\Phi}{\delta \bv} - \frac{d}{dt}\left(\frac{\delta\Phi}{\delta \dot{\bv}}\right) +\cdots\right]_{\bv=\dot{\br},\dot{\bv}=\ddot{\br},\cdots}
    =0.
    \label{eq:EL_eq_2}
\end{align}
The above pair of equations are just Eqs.~(2.3) and (2.4) from Ref.~\onlinecite{edwards1974theory}.
For simplicity, we now revert to the notation used for a single particle around Eqs.~\eqref{eq:5} and \eqref{eq:6}.
As always, ${\cal L}=K-U$ and so Eq.~\eqref{eq:EL_eq_2} becomes
\begin{equation}\label{eq:EL_3}
    -\frac{\partial U}{\partial \br}-\frac{d}{dt}\left(\frac{\partial K}{\partial \dot{\br}}\right) - \frac{\partial \Phi}{\partial \bv} =0.
\end{equation}
The first term above can be rewritten as $-\frac{\partial \dot{U}}{\partial \dot{\br}}$, which comes from noting that $\dot{U}=\frac{\partial U}{\partial \br}\dot{\br}$ (chain rule) and so $\frac{\partial \dot{U}}{\partial \dot{\br}}=\frac{\partial U}{\partial \br}$.
Using this result in Eq.~\eqref{eq:EL_3} then allows us to rewrite it as
\begin{equation}\label{eq:EL_4}
    \frac{\partial (\dot{U}+\Phi)}{\partial \bv}=-\frac{d}{dt}\left(\frac{\partial K}{\partial \bv}\right),
\end{equation}
which is just a way of writing Eq.~\eqref{eq:4}.

{\em Concluding remarks:}
In this paper we have postulated that Eq.~\eqref{eq:4} should replace the OVP in Eq.~\eqref{eq:2} whenever inertial effects are expected to be just as important as dissipative/frictional effects.
To support this, we have demonstrated that Eq.~\eqref{eq:4} is a valid and useful way to formulate the dynamical equations for colloidal particles suspended in a fluid, both in the one-particle case and for a fluid of interacting colloidal particles.
Thus, we have provided the generalisation of PFT \cite{schmidt2022power} to the isothermal under-damped Brownian dynamics case.

Thinking broadly, the OVP in Eq.~\eqref{eq:2} has turned out to be quite general and has been applied successfully to a wide range of systems where friction or viscous dissipation dominates the dynamics~\cite{doi2011onsager, doi2013soft, doi2019application} (i.e.\ not just colloidal fluids and droplets on surfaces).
For example, the OVP/PFT has been applied to active matter and biological systems.\cite{krinninger2019power, hermann2019phase, liu2026entropy, xu2026onsager}
We expect the same success for Eq.~\eqref{eq:4} whenever inertial effects start to become relevant.

To build on the ideas presented here, we see several avenues for fruitful future work. The first direction concerns the foundations of Eq.~\eqref{eq:4} and determining in greater detail the implicit assumptions and limitations.
In the literature, there are various theories of `analytical thermodynamics',\cite{podio2023analytical} the GENERIC formulation of fluid dynamics, \cite{grmela1997dynamics, grmela2026rheological} applications of the Herglotz variational principle to dissipative field theories \cite{lazo2018action, gaset2024herglotz} and the way of incorporating inertial effects developed in Ref.~\onlinecite{yasuda2026covariant}.
Working through these other approaches to see if and how they connect to the formulation in Eq.~\eqref{eq:4} should be extremely fruitful.
This will build connections to other areas, where ideas may exist that are new to the soft matter areas discussed here.
By addressing these questions and investigating these possible connections, this should allow to determine how widely applicable Eq.~\eqref{eq:4} is.

A powerful coarse-grained application/extension of the general OVP-based \eqref{eq:2} approach is to parametrise the system of interest with just a few relevant variables, resulting in simple (ordinary differential equation) models that are easy to solve numerically.
For example, for droplets on surfaces, instead of treating the full profile $h({\bf r},t)$, one can consider just the dynamics of the droplet radius and maximum height.
This simplified approach has successfully incorporated/captured evaporation, solute deposition (coffee ring effect), contact line motion, droplet interactions and more.\cite{toth2026variational, man2017vapor, wu2018multi, wu2019drying, wu2021contact, yang2021deposition}
Future work to extend and build on these approaches, by incorporating inertial effects via Eq.~\eqref{eq:4}, will allow us to determine when and how inertia changes the dynamics.
This will enable us to predict e.g.\ the onset of damped-oscillatory dynamics of liquid bridges suspended between parallel rod-shaped electrodes.\cite{bokanyi2026}

Finally, we expect our generalisation of the OVP and PFT to the under-damped regime to be especially valuable for systems that are driven to be far-from-equilibrium, since this is where inertia and friction/dissipation can become of equal significance, due to the driving.

{\em Acknowledgements:} We gratefully acknowledge useful discussions and feedback from Achilleas Lazarides, David Sibley, Tapio Ala-Nissila and Uwe Thiele.


%

\end{document}